%% file: ms.tex
\documentclass{emulateapj}
\usepackage{natbib}
\usepackage{amsmath,amsfonts,amssymb}
\usepackage{graphicx}
\usepackage[normalem]{ulem}

\newcommand\rchisq{\chi^2_\nu}
\newcommand\srchisq{{(\chi^2_\nu)^{1/2}}}

\newcommand\Mj{{\rm M_{J}}}

\newcommand\Ms{{\rm M_{\odot}}}
\newcommand\Rs{{\rm R_{\odot}}}

\newcommand\mps{{\rm m\,s^{-1}}}

\newcommand\kmps{{\rm km\,s^{-1}}}

\newcommand\msini{m \sin i}
\newcommand\vsini{{\it v}_{rot} \sin i_{\star}}
\newcommand\psini{P_{rot} (\sin i_{\star})^{-1}}
\newcommand\ii{\rm I_{2}}

\def\url#1{\texttt{#1}}

\shorttitle{Planetary-mass companions to the K-giants BD+15~2940 and HD~233604.}
\shortauthors{Nowak et al.}

\begin{document}

\title{BD+15~2940 and HD~233604: two giants with planets close to the engulfment zone}
\author{G. Nowak\altaffilmark{1}, A. Niedzielski\altaffilmark{1}, A. Wolszczan\altaffilmark{2,3}, M. Adam\'ow\altaffilmark{1}, G. Maciejewski\altaffilmark{1}}
\altaffiltext{1}{Toru\'n Center for Astronomy, Nicolaus Copernicus University, ul. Gagarina 11, 87-100 Toru\'n, Poland, grzegorz.nowak@astri.umk.pl, andrzej.niedzielski@astri.umk.pl, monika.adamow@astri.umk.pl, gracjan.maciejewski@astri.umk.pl}
\altaffiltext{2}{Department of Astronomy and Astrophysics, the Pennsylvania State University, 525 Davey Laboratory, University Park, PA 16802, alex@astro.psu.edu}
\altaffiltext{3}{Center for Exoplanets and Habitable Worlds, the Pennsylvania State University, 525 Davey Laboratory, University Park, PA 16802}

\begin{abstract}
We report the discovery of planetary-mass companions to two red giants by the ongoing Penn State - Toru\' n Planet Search (PTPS) conducted with the 9.2-m Hobby-Eberly Telescope. The 1.1~$\Ms$ K0-giant, BD+15~2940, has a 1.1~$\Mj$ minimum mass companion orbiting the star at a 137.5-day period in a 0.54~AU orbit what makes it the closest - in planet around a giant and possible subject of engulfment as the consequence of stellar evolution. HD~233604, a 1.5~$\Ms$ K5-giant, is orbited by a 6.6~$\Mj$ minimum mass planet which has a period of 192 days and a semi-major axis of only 0.75~AU making it one of the least distant planets to a giant star. The chemical composition analysis of HD~233604 reveals a relatively high {$^7$Li} abundance which may be a sign of its early evolutionary stage or recent engulfment of another planet in the system.

We also present independent detections of planetary-mass companions to HD~209458 and HD~88133, and stellar activity-induced RV variations in HD~166435, as part of the discussion of the observing and data analysis methods used in the PTPS project.
\end{abstract}

\keywords{planetary systems-stars: individual (BD+15~2940, HD~233604, HD~209458, HD~88133, HD~166435)}

\section{Introduction}
A precision of the radial velocity (RV) technique, which is the most prolific ground based method of planet detection, depends on the presence of a large number of narrow spectral lines in the stellar spectra. This limits its efficiency to slowly rotating, cool stars of late-F to M spectral types which also sets an upper limit of $M_{\star} \lesssim 1.5 M_{\odot}$ to the masses of Main Sequence (MS) target stars that may be hosts to planetary companions. Fortunately, as these stars evolve off the MS and become red giants, they cool and slow down their rotation which makes them suitable again for precision RV measurements. Despite the fact that the precision of this technique is seriously affected by stellar activity \citep{2001AaA...379..279Q} which is particularly true in the case of giant stars \citep{2003AJ....125..293C,2004A&A...421..241S}, it remains most efficient in searches for planets around A-F stars with masses as high as 5~$\Ms$. This offers an excellent opportunity to improve our understanding of the population of planets around evolved stars that are significantly more massive than the Sun.

Searches for planets around red giants also represent an essential extension of the traditional surveys, because each detection of a planetary system around a sufficiently evolved giant provides a snapshot of the changes in its dynamical configuration powered by the evolution of the parent star. In principle, this makes it possible to track the evolution of planetary systems back to their origins \citep{2012ApJ...754L..15A} and to extrapolate it beyond the red giant stage \citep{2007ApJ...661.1192V,2012ApJ...761..121M}. 

Over a dozen stars with planetary-mass companions have already been discovered by the Penn State-Toru\'n Centre for Astronomy Planet Search (PTPS), whose primary, long-term goal is to improve our understanding of the evolution of planetary systems around aging stars \citep{2007ApJ...669.1354N,2009ApJ..693..276N,2009ApJ...707..768N,2012ApJ...745...28G,2012ApJ...756...53G}. Our findings are part of the growing body of discoveries by surveys such as the McDonald Observatory Planet Search \citep{1993ASPC...36..267C,1993ApJ...413..339H}, Okayama Planet Search \citep{2003ASPC..294...51S}, Tautenburg Planet Search \citep{2005A&A...437..743H}, Lick K-giant Survey \citep{2002ApJ...576..478F}, ESO FEROS planet search \citep{2003A&A...397.1151S}, Retired A Stars and Their Companions \citep{2007ApJ...665..785J}, CORALIE \& HARPS search \citep{2007A&A...472..657L}, Boyunsen Planet Search \citep{2011AIPC.1331..122O}, and several others.

The existing population of planets around evolved intermediate-mass stars, uncovered by these surveys, begins to show its distinction from that of the solar type MS dwarfs. For example, the planet frequency-stellar metallicity correlation observed for MS dwarfs \citep{2005ApJ...622.1102F} appears to be weaker in the case of planets around giants \citep{2007A&A...473..979P,2007A&A...475.1003H,2010ApJ...725..721G}, and, compared to dwarfs, the frequency of intermediate-mass stars with planets seems to be higher \citep{2007A&A...472..657L,2010PASP..122..905J,2011ApJS..197...26J}. Moreover, the intermediate mass stars prefer to have more massive planets and are deficient in low mass ($\leq$ 1.5~$\Mj$) companions \citep{2007ApJ...661..527S,2009ApJ...707..768N,2010ApJ...709..396B}. Also intriguing is the growing population of brown dwarfs orbiting these stars \citep[e.g.][]{2008ApJ...672..553L,2009ApJ...707..768N,2011ESS.....2.2008Q}.

The most dramatic evidence for the dynamical changes in planetary systems around evolved stars has been provided by the observed absence of planets in orbits with semi-major axes smaller than 0.78~AU \citep{2007ApJ...665..785J}. This particular property of planets around red giants detected so far seems to be an imprint of stellar evolution on planetary systems architectures. Namely, the planets at small orbital separations will tidally interact with their host stars already during their MS lifetime, and some of them will likely end up spiraling into their envelopes \citep[see e.g.][]{2009ApJ...698.1357J}. However, the more distant planets, which have survived the MS evolution, may face the similar fate, when their parent stars ascend the RGB and their tidal influence extends a large distance beyond the stellar radius \citep{2009ApJ...705L..81V}.

A deficit of massive planets with short orbital periods was first noted by \cite{2002ApJ...568L.113Z}. The planet shortage in the separation range of 0.08-0.6~AU was explained by \cite{2003A&A...407..369U} as a signature of a transition region between two categories of planets that have experienced different migration scenarios. The gap in semi-major axis distribution for planets around M$\geq$1.2$\Ms$ stars was considered by \cite{2007ApJ...660..845B} who showed that the deficiency of close-in companions to giants may reflect the dearth of 0.1-0.6~AU companions to $M_{\star} \geq$1.2$\Ms$ F-dwarfs caused by a shorter viscous diffusion timescale of their protoplanetary disks. \cite{2009ApJ...694L.171C} has shown through a series of Monte Carlo simulations that a stellar-mass-dependent gas disk lifetime can explain a shortage of planets with $a <$ 0.5~AU around massive stars.

Paucity of planets orbiting clump giants within $q = a(1-e)$ = 0.68~AU periastron distance was first noted by \cite{2008PASJ...60..539S}. These authors singled out the planet engulfment at the tip of the RGB as the mechanism responsible for it. \cite{2009ApJ...705L..81V} have shown that the evolution of the star alone can quantitatively explain the observed absence of close-in planets around evolved stars, when tidal interaction is taken into account. They have also provided the minimum orbital radii inside of which planets would be engulfed by stars of various masses at the end of the RGB evolution. These authors also found out that the more massive the planet, the earlier it will be captured by the stellar envelope. \cite{2010MNRAS.408..631N} studied in detail 1~$\Ms$ star evolution from the ZAMS to the end of the post-MS phase and discussed the influence of tidal and mass-loss laws on the period gaps in orbits of 1, 10 and 100~$\Mj$ companions. Finally, \cite{2011ApJ...737...66K} found that minimum orbital radii are sensitive to stellar mass in the range 1.7-2.1~$\Ms$ and that all the known planets indeed orbit GK clump giants beyond these survival limits, in ageeement with the planet-engulfment hypothesis. These authors note, however, that planets around stars more massive than 2.1~$\Ms$ have semi-major axes that are significantly larger than the minimum value, and suggest that engulfment may not be the main reason for the observed lack of short-period giant planets.

In this paper, we present the discovery of a planet that orbits its parent star just inside the currently established $\sim$0.6~AU "zone of avoidance", and another one that has the orbital radius that is significantly smaller than 1~AU. We also describe in some detail the method of data analysis that we have developed over the duration of the PTPS project and the RV measurements of two test stars with known planets.

In Section 2 we describe our observing setup and give parameters of the stars discussed here. In Section 3 we present the method of data analysis and discussion of the long term stability of the HET/HRS data acquisition. The analysis of RV measurements of BD+15~2940 and HD~233604 is given in Section 4, followed by the accompanying analysis of rotation and stellar activity indicators in Section 5. Finally, our results are summarized and further discussed in Section 6.

\section{Observations and stellar parameters}
Observations presented in this paper were made with the Hobby-Eberly Telescope (HET) \citep{1998SPIE.3352...34R} equipped with the High Resolution Spectrograph (HRS) \citep{1998SPIE.3355..387T} in the queue scheduled mode \citep{2007PASP..119..556S}. The configuration and observing procedure employed in our program were identical to those described by \cite{2004ApJ...611L.133C}. The spectrograph, fed with the 2 arcsec fiber, was used in the R = 60,000 resolution mode with a gas cell ($\ii$) inserted into the optical path.

The spectra consisted of 46 echelle orders recorded on the ``blue'' CCD chip (407.6-592~nm) and 24 orders on the ``red'' one (602-783.8~nm). The spectral data used for RV measurements were extracted from the 17 orders, which cover the 505 to 592~nm range of the $\ii$ cell spectrum. The spectra were reduced with the dedicated IRAF/Python pipeline.

\subsection{BD+15~2940}
BD+15~2940 (HIC~78407) is a $V_{T}$ = 9.185 $\pm$ 0.024, $B-V$ = 1.006 $\pm$ 0.036 \citep{2007ASSL..350.....V} K0 \citep{1924hdhc.bookQ....C} star in Serpens. Its Hipparcos parallax $\pi$ = 1.71 $\pm$ 1.33 points to a distance of 585~pc. The star belongs to the PTPS Red Clump Giants sample \citep{2012A&A...547A..91Z}. We collected 38 spectra of BD+15~2940 over 2362 days between 2005 March 23 and 2011 September 10 (MJD 53452-55814). The signal-to-noise ratio per resolution element (S/N) in the spectra was 120-330 at 594~nm in 316-919 seconds of integration, depending on the atmospheric conditions.

\subsection{HD~233604}
HD~233604 (BD+54~1280) is $V_{T}$ = 10.406 $\pm$ 0.049 \citep{2000A&A...355L..27H}, $B-V$ = 1.007 $\pm$ 0.115 \citep{1997ESASP1200.....P}, K5 \citep{1924hdhc.bookQ....C} star in Ursa Major. It belongs to the PTPS Red Clump Giant sample and it was observed at 33 epochs over the period of 2956 days from 2004 January 11 to 2012 February 14 (MJD 53015-55971). The S/N in the spectra was 110-300 at 594~nm in 762-2100 seconds of integration, depending on the atmospheric conditions.

\subsection{HD~209458}
HD~209458 (BD+18~4917, V376~Peg) a $V_{T}$ = 7.703 $\pm$ 0.012, $B-V$ = 0.594 $\pm$ 0.015 \citep{2007ASSL..350.....V} G0 dwarf \citep{2001MNRAS.328...45M}, is known \citep{2000ApJ...529L..41H,2000ApJ...532L..55M} as a host of a transiting \citep{2000ApJ...529L..45C}, $\msini$ = 0.689 $\pm$ 0.024~$\Mj$ planet at $a$ = 0.04723 $\pm$ 0.00079~AU orbit \citep{2006ApJ...646..505B} ($P$ = 3.52474859 $\pm$ 3.8$\times$10$^{-7}$ d \citep{2007ApJ...655..564K}) with RV semi-amplitude of $K$ = 84.67 $\pm$ 0.7~$\mps$ \citep{2008ApJ...677.1324T}. This star is one of the PTPS test stars, which are known hosts to planetary systems that we have been monitoring as part of our long-term observing program. We have observed HD~209458 between 2004 July 11 and 2011 November 25 over 2693 days or 7.4 years (MJD 53197-55890) and collected 29 epochs of RV measurements. The exposure times varied between 94 and 608 seconds depending on weather conditions, typically achieving the S/N of about 200.

\subsection{HD~88133}
HD~88133 (BD+18~2326), a $V_{T}$ = 8.094 $\pm$ 0.013, $B-V$ = 0.810 $\pm$ 0.015 \citep{2007ASSL..350.....V} G5 dwarf \citep{1924hdhc.bookQ....C}, is another PTPS test star, a well-studied short-period host of a $\msini$ = 0.299 $\pm$ 0.027~$\Mj$ planet in $a$ = 0.04717 $\pm$ 0.00079~AU ($P$ = 3.41587 $\pm$ 0.00059 d) with the RV semi-amplitude of $K$ = 36.1 $\pm$ 3~$\mps$ \citep{2005ApJ...620..481F,2006ApJ...646..505B}. It was occasionally observed between 2006 January 23 and 2012 February 20 (MJD 53758-55977). 23 epochs of observations were gathered with S/N of 110-330 achieved in 141-800 seconds exposures, depending on actual weather conditions.

\subsection{HD~166435}
HD~166435 (BD+29~3190), a $V_{T}$ = 6.901 $\pm$ 0.010, $B-V$ = 0.633 $\pm$ 0.006 \citep{2007ASSL..350.....V} G1 sub-dwarf \citep{2007AJ....133.2524W}, has been known to exhibit a pronounced stellar activity due to spots that manifests themselves as the measured RV variations induced by the line bisector variability of an amplitude of $\sim$135~$\mps$ \citep{2001AaA...379..279Q}. As one of the PTPS test stars, it was observed between 2011 February 19 and 2011 September 22 (MJD 55611-55826) at 23 epochs. The observations resulted in S/N of 290-540 obtained with 70-700 seconds exposures, depending on actual weather conditions.

\subsection{Properties of the stars}
The atmospheric parameters ($T_{eff}$, $\log g$, [Fe/H], $v_{micro}$) of BD+15~2940 and HD~233604 were taken from \cite{2012A&A...547A..91Z}, who have estimated their values using the method of \cite{2002PASJ...54..451T,2005PASJ...57...27T}. With these values and the luminosities estimated from available data, stellar masses and ages were derived by fitting the ensemble of parameters characterizing the star ($\log L_{\star}/L_{\odot}$, $\log T_{eff}$, $\log g$, and metallicity) to the evolutionary tracks of \cite{2000A&AS..141..371G}. The stellar radii were estimated using the derived masses and $\log g$ from the spectroscopic analysis as described in \cite{2012A&A...547A..91Z}. The stellar rotation velocities of BD+15~2940 and HD~233604 were taken from Nowak et al. (2013, in preparation), who have estimated their values using the cross-correlation technique and calibrations of \cite{1997PASP..109..514F} and \cite{2007A&A...475.1003H}. The stellar parameters are summarized in Table~\ref{tab-01}.

For HD~209458 and HD~88133 we adpoted the stellar parameters determined in \cite{2005ApJS..159..141V} and \cite{2005ApJ...620..481F}, respectively.

\subsubsection{Li abundance}
Lithium abundance analysis was performed with the Spectroscopy Made Easy package \cite[SME,][]{1996A&AS..118..595V}, which requires as an input a set of lines from the Vienna Atomic Line Database \cite[VALD,][]{1999A&AS..138..119K} identified in the spectrum, the stellar data (including RVs), and the instrumental profile. Instead of the [M/H] ratio, also needed by the SME, the [Fe/H] values from \cite{2012A&A...547A..91Z} were used as the best approximation. The 6695-6725~{\AA} range, which covers both the $^7$Li line at 6708~{\AA} and a set of Al, Ti, Si and Ca spectral lines, was modeled with the SME using the RV, rotation velocity, and macroturbulence velocity as parameters. The uncertainty of the lithium abundance estimates was derived from the rms scatter of the SME fits to all the spectra available for a given star. In addition, the analysis of lithium abundances requires a correction for non-LTE effects, especially in the case of the lithium line at 6708~{\AA} considered here, which is affected by several non-LTE processes \cite[][and references therein]{1994A&A...288..860C}. We have applied the non-LTE correction from \cite{2009A&A...503..541L} to account for that effect. We have derived the values of A(Li) of 1.595 $\pm$ 0.093, 3.104 $\pm$ 0.056, and 1.275 $\pm$ 0.188 for HD~88133, HD~209458 and HD~166435, respectively. Given their evolutionary status, the Li content of these stars is not abnormal. For BD+15~2940 we have measured A(Li) = 0.247 $\pm$ 0.132, while for HD~233604 A(Li) = 1.400 $\pm$ 0.042, which makes it a Li-reach giant.

\section{Data analysis}
\subsection{Radial velocity and line bisector measurements}
As the HRS is a general purpose spectrograph, which is neither temperature nor pressure controlled, calibration of RV measurements with this instrument is best accomplished with the $\ii$ cell technique. We have improved the efficiency of the entire data analysis process by combining this method \citep{1992PASP..104..270M,1996PASP..108..500B} with that of cross-correlation \citep{1995IAUS..167..221Q,2002A&A...388..632P} for the purpose of the respective RV, line bisector span (BS) and line bisector curvature (BC) measurements in the final version of the ALICE code.

A detailed implementation of the above approach is described in \cite{2013PhDT..........N}. Briefly, we used the $\ii$ spectrum imprinted on the HRS flat-field spectrum to correct the initial ThAr lamp-derived wavelength scale and to determine the instrumental profile (IP) modeled as a sum of five Gaussian profiles. The wavelength scale was assumed to be linear over the 96-pixel segments that the spectra were divided into. The high S/N stellar template spectrum was then smoothed using optimal Wiener filtering \citep{1975tads.book.....R}, deconvolved with the model IP using \cite{1984dwas.book.....J} method, resampled to the corrected wavelength scale, and used as a model for RV determination from a remaining shift between the model and the stellar spectrum observed at a given epoch. The modeling process employed the non-linear least squares Levenberg-Marquardt (L-M) method \citep{1992nrfa.book.....P}. The RV for each epoch was derived as a mean value of 391 independent measurements from the 17 usable echelle orders, after rejecting those with outlying radial velocities (more than 3$\sigma$ off the mean value) and those obtained for segments without stellar spectral lines, identified in course of cleaning the combined stellar and iodine spectra from the $\ii$ lines. The RV uncertainty for each epoch was calculated as the error of the mean value ($=rms/\sqrt{n_{seg}}$, where $n_{seg}$  is the number of actually used segments).

A computation of the cross-correlation functions (CCF) for the purpose of BS/BC determination was carried out as part of the above RV measurement procedure. To this end, the combined stellar and iodine spectra were first cleaned of the $\ii$ lines by dividing them by the corresponding iodine spectra imprinted in a flat-field, and then cross-correlated with a binary CCF mask. Typically the cleaned spectra and stellar template agree within 1\%, consistent with the S/N ratio of 100. The mask was constructed from a synthetic K2 star spectrum of \cite{1993KurCD..13.....K} Atlas 9 and contained about 300 lines.

The CCFs were computed in all the 17 HRS orders used for RV measurements over a wide, $\pm$ 30~$\kmps$ range, after correcting the spectra for their absolute radial velocity and shifting them to the Solar System barycenter \citep{1980A&AS...41....1S}. Finally, the 17 CCFs were added together in the wavelength scale to form the integrated CCF used in further analysis. The absolute RVs at a given epoch of observations were measured from the position of a maximum of the CCF, while the CCF line bisectors and curvatures, defined in the usual way, were calculated using 3 reference ranges of (0.1~--~0.25), (0.375~--~0.525) and (0.65~--~0.8) of the central CCF intensity. The uncertainties of BS measurements were estimated with the formulae of \cite{2005A&A...442..775M}.

\subsection{Modeling of radial velocity variations}
We fitted Keplerian orbits to the observed RV variations using a hybrid approach \citep{2003ApJ...594.1019G,2006A&A...449.1219G,2007ApJ...657..546G}, which combines the global search for a plausible range of orbital parameters with the aid of the genetic algorithm (GA), followed by a non-linear least squares or a downhill simplex fit to data to quickly converge to the preliminarily constrained $\chi^2$ minimum. 

In our implementation of this approach, once a periodic signal in a RV time series was identified in its Lomb-Scargle (L-S) periodogram \citep{1976Ap&SS..39..447L,1982ApJ...263..835S}, we searched for possible orbital solutions over a wide range of parameters with the PIKAIA based GA code \citep{1995ApJS..101..309C}. The GA semi-global search usually identified a narrow parameter range of the search space, which was then explored using the L-M algorithm to locate the best-fit Keplerian solution. In the case of a single planet the RV scrambling method \citep{2004ApJ...617..580B} was used to assess the false alert probability (FAP) of the final orbital solution. If a more complicated system was indicated by a presence of two or more periodicities in the RV data, the scrambled residua were used to quantify the goodness of the final solution. Uncertainties of the best-fit orbital parameters were estimated from either the covariance matrix of the L-M fits or from the scrambled residuals \citep{2005ApJ...619..570M}.

\subsection{Monitoring the long term stability of the HET/HRS data acquisition}
Along with the target stars, we have been monitoring three stars of known RV behavior, HD~209458, HD~88133, and HD~166435, to track any possible changes in the data acquisition system that could affect the precision of RV measurements. 

\begin{figure}
\centerline{\includegraphics[angle=0,scale=2.5]{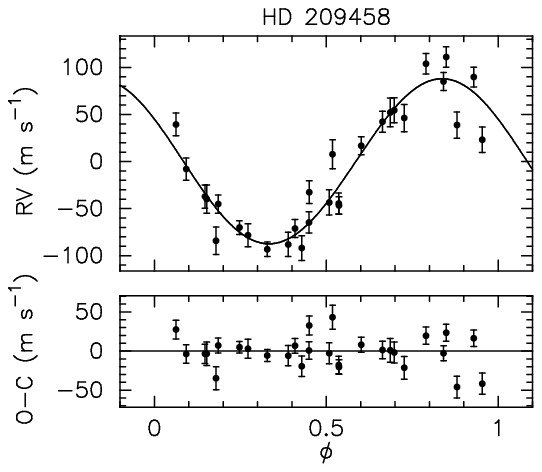}}
\caption{Radial velocities of HD~209458 plotted as a function of orbital phase for the solution detailed in Table~\ref{tab-05}. The measurements taken at or near the transits have been removed.\label{fig-01}}
\end{figure}

The RV and BS measurements for HD~209458 are listed in Table~\ref{tab-02}. The long-term RV uncertainty of 12~$\mps$ for this star was dominated by its high $T_{eff}$, which resulted in a correspondingly low number of spectral lines that could be used in RV measurements. The uncertainty of the BS estimates amounted to 38.9~$\mps$. The L-S periodogram of the observed RV variations revealed a clear, single period of 3.52 days and no trace of any long-term trend. The best fit Keplerian model of these data is shown in Figure~\ref{fig-01} and the corresponding orbital parameters are given in Table~\ref{tab-05}. Within uncertainties, our results are identical to those of \cite{2006ApJ...646..505B}.

\begin{figure}
\centerline{\includegraphics[angle=0,scale=2.5]{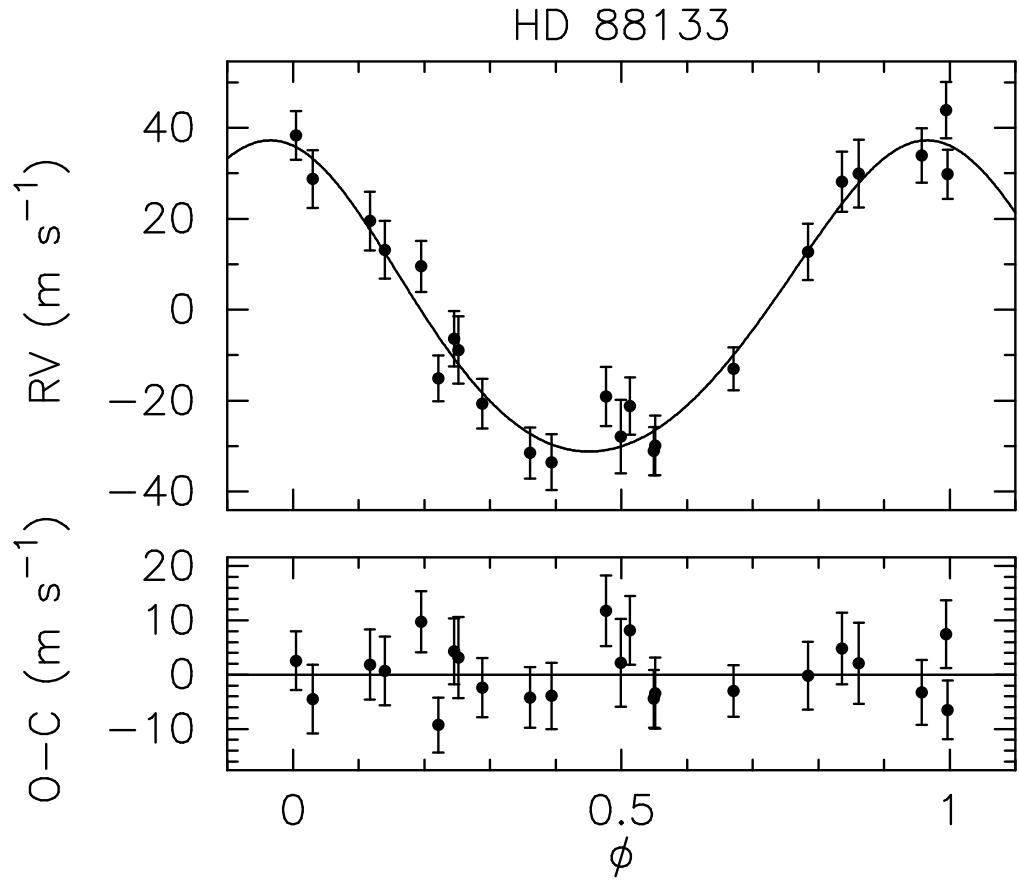}}
\caption{Radial velocities of HD~88133 plotted as a function of orbital phase for the solution detailed in Table~\ref{tab-05}.\label{fig-02}}
\end{figure}

The RV and BS measurements for HD~88133 are listed in Table~\ref{tab-03}. In the case of this star, the 6.2~$\mps$ RV uncertainty was much lower than for HD~209458, and the respective uncertainty of BS measurements was 16.5~$\mps$. The L-S periodogram of the RV time series exhibited a highly significant signal at a 3.41-day period, and, as in the case of HD~209458, no long-period trend. The best-fit Keplerian orbit, folded at the orbital period is presented in Figure~\ref{fig-02} and the orbital parameters are listed in Table~\ref{tab-05}. Once again, within uncertainties, our results are identical to those of \cite{2006ApJ...646..505B}.

\begin{figure}
\centerline{\includegraphics[angle=0,scale=2.0]{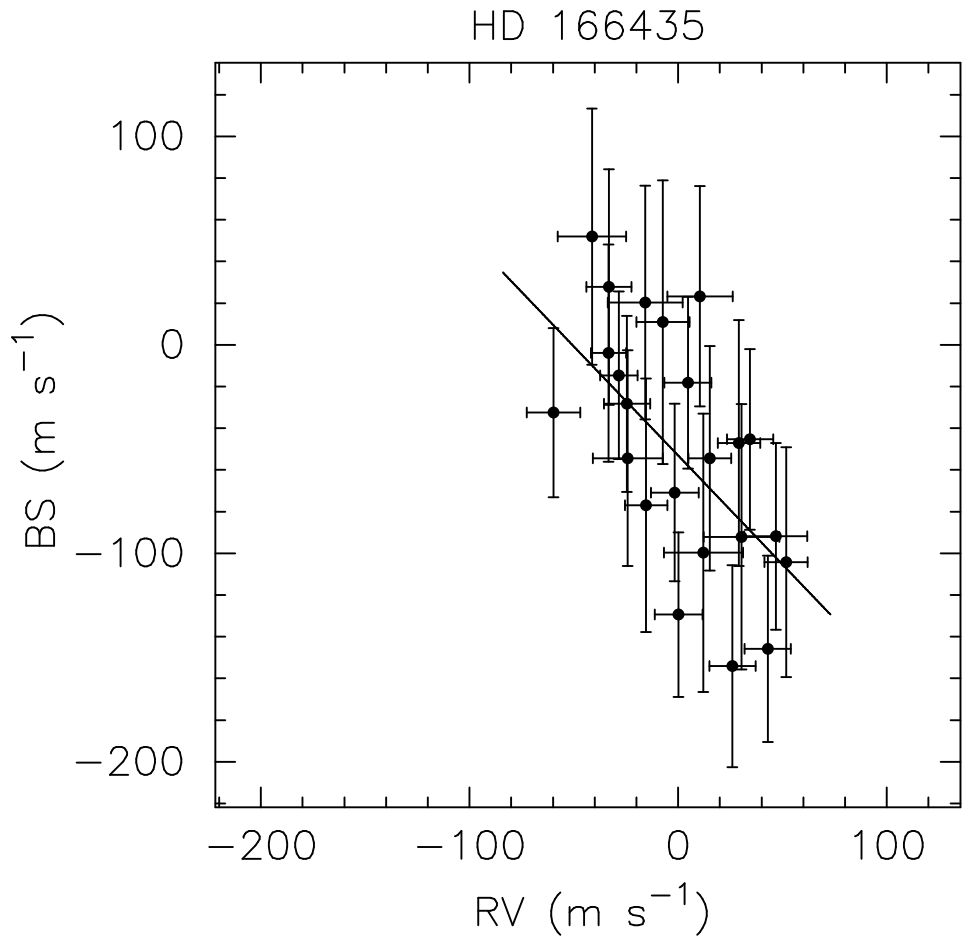}}
\caption{Radial velocity versus bisector span of the CCF profile of HD~166435. The line represents the best linear fit to the data.\label{fig-03}}
\end{figure}

Finally, despite the fact that the 51.5~$\mps$ uncertainty of the BS measurements for HD~166435 (Table~\ref{tab-04}) was much larger than the respective RV uncertainty of 12.7~$\mps$, the correlation between the two quantities was obvious ($r$ = 0.63, while the critical value of the Pearson correlation coefficient at the confidence level of 0.01 is r$_{21,0.01}$ = 0.52, Figure~\ref{fig-03}), in agreement with the results reported by \cite{2001AaA...379..279Q} for this star. This confirms the reliability of our BS measurements. Unfortunately, given the poor phase coverage of our observations we were not in position to confirm the periodic character of the RV and BS variations that mimic a planet. 

Overall, the results from RV and BS monitoring of the three stars discussed above demonstrate a satisfying long-term stability of our data acquisition scheme.

\section{Results}
\subsection{BD+15~2940}
\begin{figure}
\plottwo{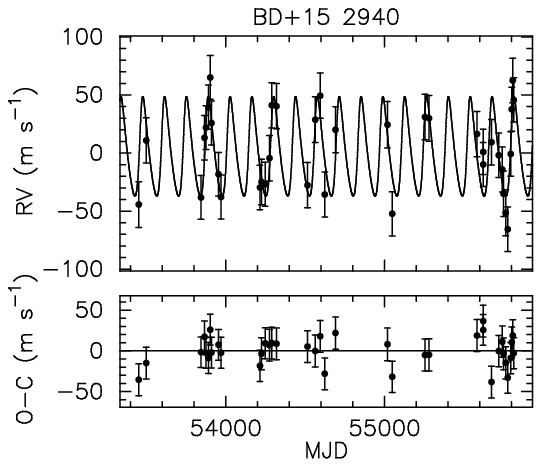}{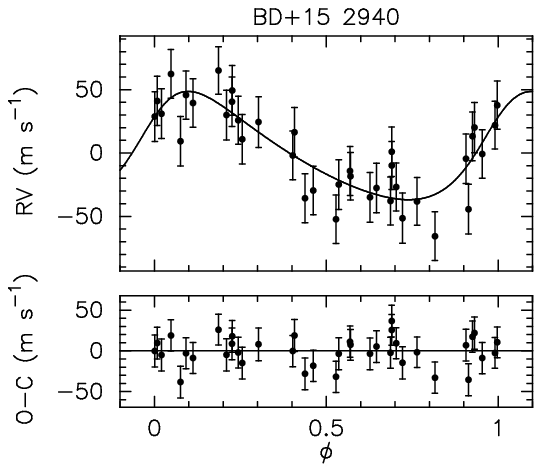}
\caption{Left: radial velocities (filled circles), the best fit  Keplerian model detailed in Table~\ref{tab-08} (full line) and the post-fit residuals for BD+15~2940. Right: the same as a function of orbital phase.\label{fig-04}}
\end{figure}

RV measurements of BD+15~2940 made at 38 epochs over a period of almost 6.5 years are listed in Table~\ref{tab-06} and shown in Figure~\ref{fig-04}. The estimated RV uncertainty for this star was $\sigma_{RV}$ = 7~$\mps$ with the expected amplitude of solar-type oscillations of $K_{osc}$ = 22~$\mps$ \citep{1995A&A...293...87K}. The star exhibits RV variations with an amplitude of 65~$\mps$ which is $\sim$3 times the expected RV uncertainty. 

\begin{figure}
\centerline{\includegraphics[angle=0,scale=2.0]{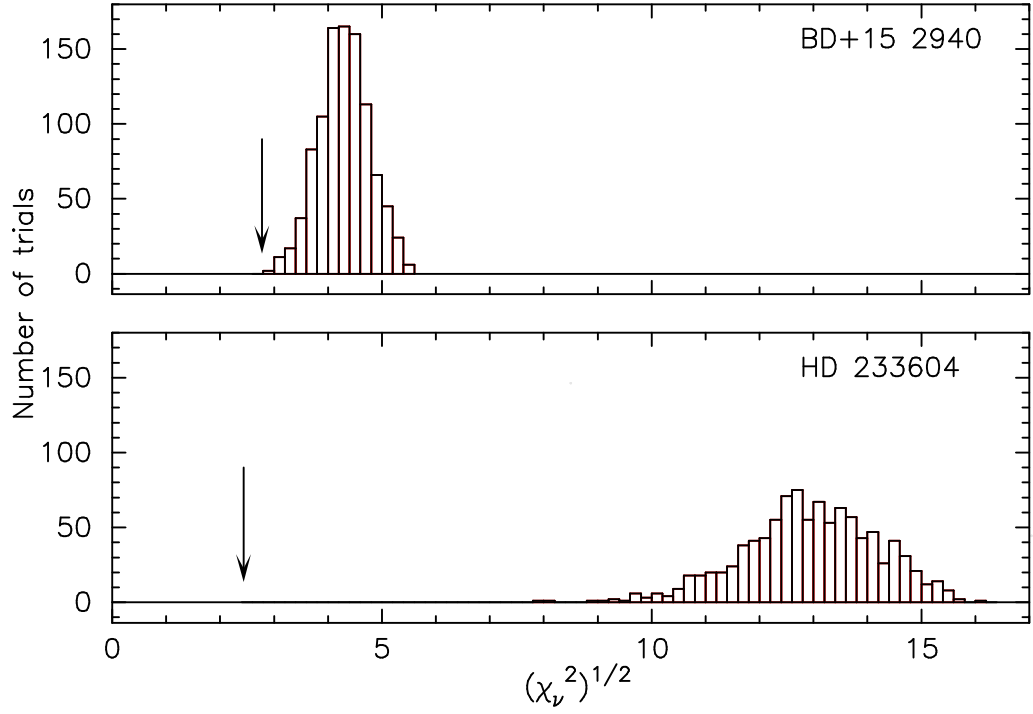}}
\caption{Histograms of the values of $\srchisq$ obtained from the fits of the Keplerian models to scrambled RVs used to estimate the FAPs for the two stars discussed in the text. In each case, 1000 sets of scrambled RVs have been generated. Vertical arrows point to the respective $\srchisq$ values derived from fits to unscrambled velocities. In both cases the FAP values are less than 0.1\%.\label{fig-08}}
\end{figure}

The L-S periodogram of the RV variations for this star revealed a single period of 137.5 days (Figure~\ref{fig-06}~a) and no signature of any low frequency component. The best-fit Keplerian orbit is shown in Figure~\ref{fig-04} and its parameters are listed in Table~\ref{tab-08}. When an additional scatter $\sigma_{+}$ of 18~$\mps$ determined from this fit was added in quadrature to the original RV uncertainties, it resulted in a $\rchisq$ value of 0.989. Note that this increase in the long-term RV uncertainty is very close to the above $K_{osc}$ estimate and the actual jitter derived as $\sigma_{j}=\sqrt{\sigma^{2}_{O-C} - \sigma^{2}_{RV}}$. The FAP$<$0.001 for this orbital solution was estimated by performing 1000 Keplerian fits to the scrambled data points as shown in Figure~\ref{fig-08}.

Given the estimated stellar mass of 1.1~$\Ms$ and the orbital period of $P$ = 137.5 days, the semi-major axis of the orbit of the planet is 0.54~AU, and the RV semi-amplitude of $K$ = 43~$\mps$ gives its minimum mass of 1.1~$\Mj$. The moderately eccentric orbit of $e$ = 0.26 $\pm$ 0.10 makes the planet's distance to its star vary within 0.4-0.68~AU. Therefore, the BD+15~2940 planet is the closest one to a giant star detected so far by means of the Doppler velocity method, and the second least-massive one of all planets orbiting giants within 1~AU after BD+48~738~b \citep{2012ApJ...745...28G}.

\subsection{HD~233604}
\begin{figure}
\plottwo{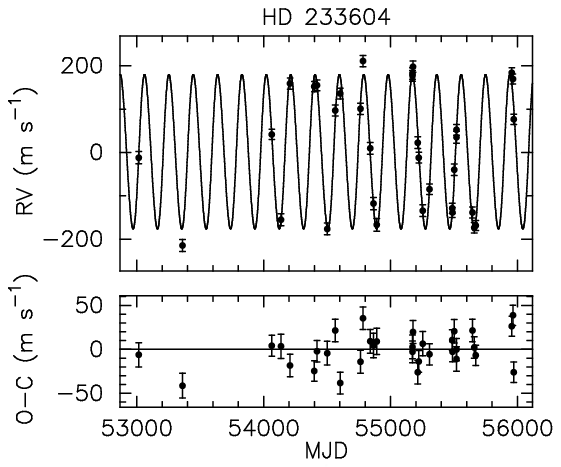}{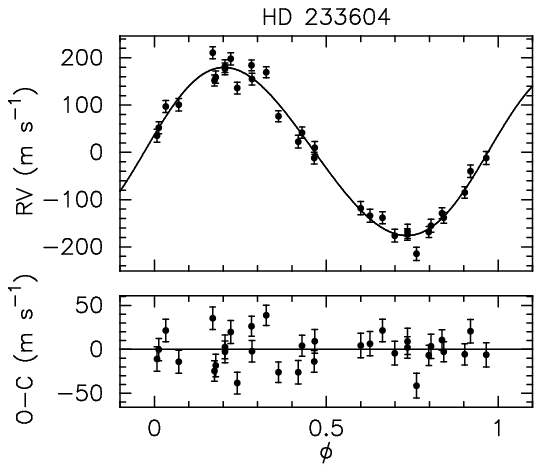}
\caption{Left: radial velocities (filled circles), the best fit  Keplerian model detailed in Table~\ref{tab-08} (full line) and the post-fit residuals for HD~233604. Right: the same as a function of orbital phase.\label{fig-05}}
\end{figure}

Radial velocities of HD~233604 (BD+54~1280) measured at 33 epochs over the period of 2956 days (over 8 years) are listed in Table~\ref{tab-07} and presented in Figure~\ref{fig-05}. The estimated RV uncertainty for this star was 9.0~$\mps$, which is exactly the same as the estimated $K_{osc}$ value. The measured RV amplitude was 213~$\mps$ or $\sim$ 17 times the expected RV uncertainty.

The dominant feature in the periodogram of these RV data is the peak at a frequency corresponding to a 192-day period. The corresponding best-fit Keplerian orbit, presented in Table~\ref{tab-08} and in Figure~\ref{fig-05}, gives the value of $\rchisq$ of 2.76. An additional scatter of $\sigma_{+}$=9~$\mps$ was added to the original errors to account for an unresolved RV variability that has apparently degraded the precision of RV measurements for this star. As in the previous case, the FAP of $<$0.001 (Figure~\ref{fig-08}) was calculated for this best-fit model. The orbital solution for HD~233604, together with its estimated mass of $M_{\star}$ = 1.5~$\Ms$ gives a $\msini$ = 6.6~$\Mj$ companion in a 0.75~AU, nearly circular orbit with $e$ = 0.05 $\pm$ 0.03.

\section{Stellar activity analysis: line bisectors and stellar photometry.}
In order to verify that the observed RV periodicities are indeed caused by the Keplerian motion, we have thoroughly examined the existing photometry data in search for any periodic light variations, and performed a complete analysis of line bisectors and curvatures for stars discussed in this paper.

\subsection{BD+15~2940}
With the estimated rotation velocity of $\vsini$ = 2.3 $\pm$ 0.8~$\kmps$ and the radius of $R_{\star}/R_{\odot}$ = 14.7 $\pm$ 2.8 the expected upper limit to the rotation period of BD+15~2940 is 317 days, but it may range between 196 and 438 days. Therefore, it is important to consider possible sources of stellar activity that may result in the observed periodic RV signal. Such processes include stellar pulsations and a spot rotating with a star.

The p-mode oscillations can be excluded, because their expected amplitude and period are $K_{osc}$ = 22~$\mps$ and P$_{osc}$ = 0.68 d, respectively \citep{1995A&A...293...87K}. We do not resolve such the fast oscillation with our sampling that is focused on long-term RV variations. Consequently, this effect only contribute a RV-jitter to our measurements.

\begin{figure}
\centerline{\includegraphics[angle=0,scale=0.75]{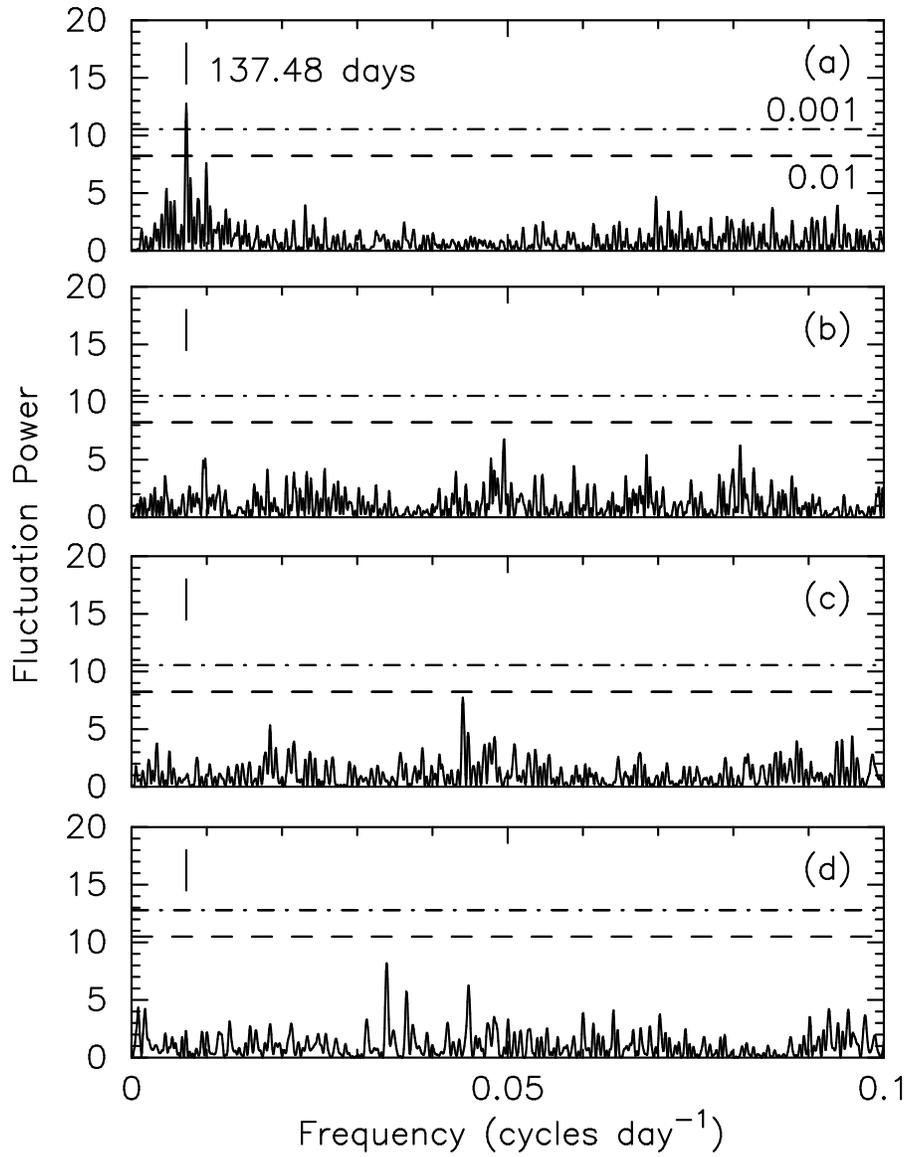}}
\caption{The L-S periodograms of (a) radial velocities (b) RV residua, (c) bisector span and (d) ASAS \citep{1997AcA....47..467P} photometry of BD+15~2940. The levels of FAP=1.0\% and 0.1\% are shown.\label{fig-06}}
\end{figure}

The mean values of the BS and the BC for this star are -5 $\pm$ 23~$\mps$, and -20 $\pm$ 32~$\mps$, respectively, with the corresponding uncertainties of 18~$\mps$ and 24~$\mps$. The BS and BC amplitudes are 45~$\mps$ and 75~$\mps$. Neither BS (Figure~\ref{fig-06}~c) nor BC show any periodic signal above the f$_{n}$ = 0.01 level. Pearson's correlation coefficients for RV and BS and BC are 0.12 and 0.17, respectively, with the critical value at the confidence level of 0.01 is r$_{36,0.01}$ = 0.41.

Two archival photometry datasets are available for BD+15~2940. The existing Hipparcos photometry \citep{1997ESASP1200.....P} includes 65 epochs that uniformly cover the 1159 days (3.2 years) or $\sim$8.5 RV periods between MJD 47904-49063, over 12 years earlier than our RV data. No periodic signal is present in these data, either. The 362 epochs of ASAS photometry \citep{1997AcA....47..467P} collected between MJD 52685-55089 cover $\sim$6.5 years (2404 days) or $\sim$12 RV periods. These measurements, clumped in 7 groups, are contemporaneous with $\sim$70\% of our RV data. Again, no trace of a periodicity is present in these data (see Figure~\ref{fig-06}~d).

Since the ASAS photometric uncertainty of $\sigma$ = 0.015 mag is typical given the observed brightness of the star, it is unlikely that it may be a noise caused by a stellar spot. However, if we made such an assumption and postulated a rotating spot that covers 1.5\% of the stellar surface, the expected semi-amplitudes of RV and BS variations would be 61-95~$\mps$ and 12-19~$\mps$, respectively, depending on the assumed model \citep{1997ApJ...485..319S,2002AN....323..392H,2007A&A...473..983D}. Although these values are in a reasonable agreement with our RV and BS data, there is no periodicity in the ASAS data gathered over $\sim$12 RV periods, and no correlation between our RV and BS measurements. Adding to these results a moderate eccentricity of the assumed Keplerian orbit, we can practically rule out a possibility that the observed RV periodicity is caused by any of the stellar phenomena discussed above and is most likely the effect of an orbital motion of a substellar mass companion to the star.

\subsection{HD~233604}
The estimated upper limit to the rotation period of HD~233604 is 282 days and it is fairly close to the observed period of the RV variations of this star. Given the uncertainty in stellar radius and the precision of rotation velocity determination, the rotation period may range from 181 to 383 days. Obviously, it is critically important to examine the available data for a possible alternative explanation of the periodic RV signal.

Due to the length of the period of the observed RV variations and their high amplitude we can confidently reject p-mode oscillations as a cause of the RV periodicity. The most likely alternative is a spot on the stellar surface moving with the star at its rotation period. Such a spot should be detectable in both the photometric and the line profile variations. 

\begin{figure}
\centerline{\includegraphics[angle=0,scale=0.75]{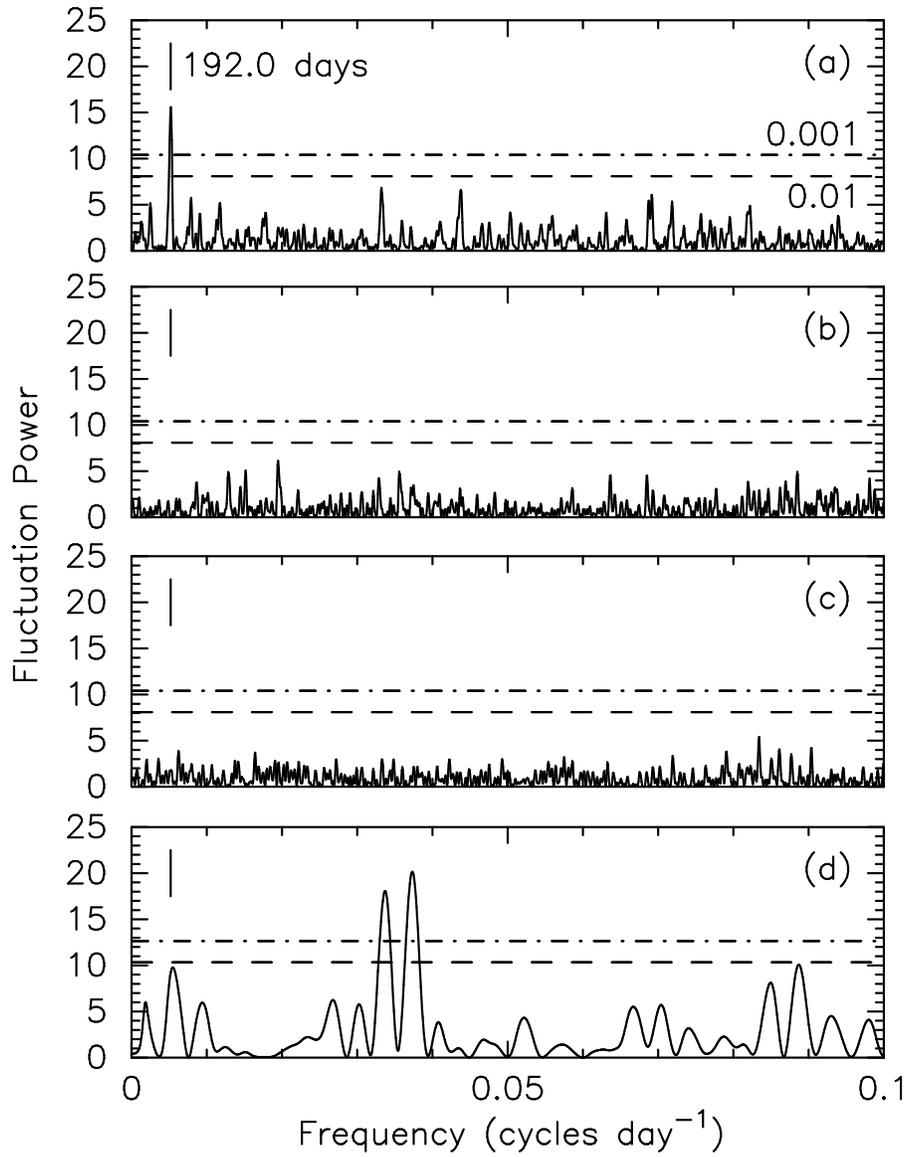}}
\caption{The L-S periodograms of (a) radial velocities and their residua (b), (c) bisector span and (d) NSVS \citep{2004AJ....127.2436W} photometry of HD~233604. The levels of FAP=1.0\% and 0.1\% are shown.\label{fig-07}}
\end{figure}

The amplitudes and mean values of the BS and BC measurements for this star are 65~$\mps$ and 9.6 $\pm$ 30.1~$\mps$, and 52~$\mps$ and -9.7 $\pm$ 32.3~$\mps$, respectively, and the corresponding uncertainties are 22 and 30 $\mps$. Neither BS (Figure~\ref{fig-07}~c) nor BC variations show any evidence for a periodicity in the computed LS periodograms. Similarly, no correlations between the RV and the BS and BC variability were found. The respective Pearson's correlation coefficients are $r$ = 0.34 for the BS and $r$ = 0.20 for the BC (with critical value of $r_{36, 0.01}$ = 0.44 for the confidence level of 0.01). 

In the case of this star, long-term photometric time series are available from the NSVS \citep{2004AJ....127.2436W} and the WASP \citep{2006PASP..118.1407P}. SWASP photometry consists of 2767 measurements between MJD 54189 and 54575, which are contemporaneous with our spectroscopic observations but cover only a small part of their time baseline. In fact, most of the available epochs of photometry fall within 175 days between MJD 54400 through MJD 54575. The NSVS photometry is of a better quality and consists of 309 measurements made between MJD 51272 and MJD 51633, which amounts to 361 days of data collected about 4 years earlier than our HET observations. These data are clustered in two groups, each covering a fraction of the orbital period. 

The LS analysis reveals no periodic signal in SWASP data, but four significant periodicities are found in the NSVS photometry with f$_{n}<$0.001. Two of them that are close to a 1-day period are most likely artifacts of the particular cadence of the RV observations. Two other two at 28.1 and 27.7 days are most likely caused by the residual lunar illumination of the field that was not removed in the photometric data reduction process. 

As in the case of BD+15~2940, we can make a very conservative assumption that the SWASP photometric uncertainty of $\sigma$ = 0.018 mag is caused by a spot, covering 1.8\% of the stellar surface. This gives the expected semi-amplitude of RV and BS variations of 70-111~$\mps$ and 13-20~$\mps$, respectively, depending on the assumed model \citep{1997ApJ...485..319S,2002AN....323..392H,2007A&A...473..983D}. The expected RV variations estimated under such an unfavorable assumption are 1.6-2.5 times smaller than those observed.

Altogether, the absence of the periodicity seen in the RV data in the available photometric data, and no correlation between BC, BS and RV variations make it unlikely that they are induced by stellar activity. An additional evidence supporting the Keplerian origin of the observed periodic RV variations is provided by the fact that they are much larger than expected under the assumption that the photometric data scatter is dominated by a rotating spot on the stellar surface.

\section{Discussion}
In this paper we have presented the detection of two planetary-mass companions to two red giants in the solar to intermediate-mass range. HD~233604 hosts a $\msini$ = 6.6~$\Mj$ planet in a 0.75~AU or $\sim$15$R_{\star}$ orbit. The planet's closest approach to the star at the periastron distance of $q$ is 0.71~AU or $\sim$14 stellar radii. BD+15~2940 harbors a $\msini$ = 1.1~$\Mj$ planet in a 0.54~AU or only $\sim$7.9~$R_{\star}$ orbit which approaches the star as close as $q$ = 0.4~AU or $\sim$5.8~$R_{\star}$. Both giants belong to the Red Giant Clump and, given a large uncertainty in their luminosity determinations, their exact evolutionary stages and ages are not known. They are most likely evolving upwards the red giant branch toward the helium-flash, but their location on the horizontal branch cannot be excluded.

The case of the $M_{\star}$ = 1.5~$\Ms$ star, HD~233604, is of special interest, as this giant exhibits a relatively high Li abundance. This suggests that the star is still before the first dredge-up process \citep{1967ApJ...147..624I}. Alternatively, but less likely, it may be another case of a recent engulfment episode similar to that of BD+48~740 \citep{2012ApJ...754L..15A}. The more likely pre-dredge-up scenario would mean that the HD~233604 planetary system was not yet affected by a substantial stellar radius growth before the helium flash and the subsequent severe mass-loss. The low rotation velocity of HD~233604 additionally supports this scenario. A more detailed, ongoing chemical composition study will allow constraining the evolutionary status of this star much better.

The status of the $M_{\star}$ = 1.1~$\Ms$ star, BD+15~2940, is less certain. Its poorly determined but definitely relatively high luminosity suggests that the star is a rather evolved, He-burning object already past the episode of stellar radius expansion. 

\begin{figure}
\centerline{\includegraphics[angle=0,scale=2.0]{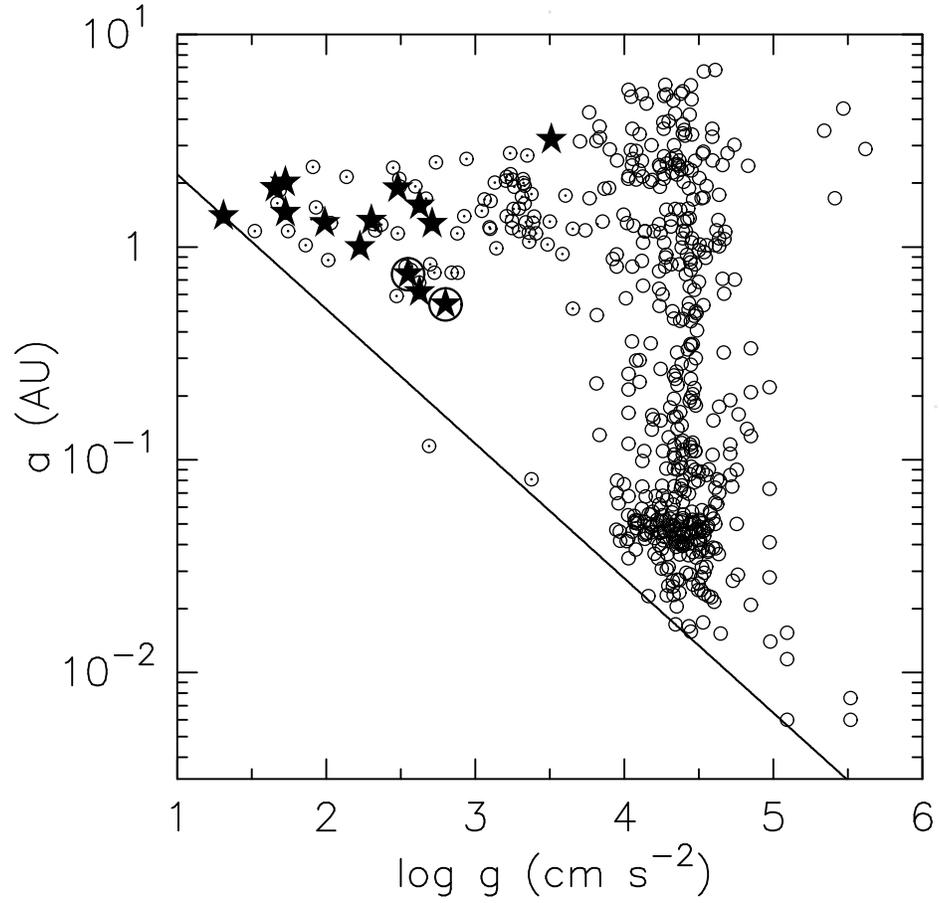}}
\caption{Semi-major axes of 778 planets as a function of $\log g$ of their host stars. The sample of 84 planets orbiting giants and subgiants with $R_{\star}$ $\geq$ 3~$\Rs$ ($\odot$) includes 12 planets discovered witin PTPS survey ($\star$). BD+15~2940~b and HD~233604~b are indicated by $\star$ with a rim.}\label{fig-09}
\end{figure}

A useful way to illustrate the occurrence of tight-orbit planets around various types of stars is to plot their semi-major-axes against the values of the stellar $\log g$. Such a plot of the current catalog of 778 planets \citep{2011A&A...532A..79S} is shown in Figure~\ref{fig-09}. Clearly, the most recent data show that there are no planets within the borderline of $\log (a[{\mathrm AU}]) = -0.633 \log g + 0.975$ (set by HD 96127 b and Kepler-10 b positions in that graph). The only exception is the intriguing system of HIP~13044 \citep{2010Sci...330.1642S}.

Planets over the whole range of orbital separations, all the way down to Hot Jupiters, can be found around MS stars with 4$< \log g \leq$5. Also, sub-giant ($\log g \sim$ 3.5) companions still cover a wide range of orbital radii, with the current lower limit of 0.0810~AU in the case of HD~102956 \citep{2010ApJ...721L.153J}, but, interestingly, there are very few planets inside 1~AU of these stars. In particular none is known within 0.08~AU$< a <$0.76~AU in accordance with the predictions of \citep{2007ApJ...660..845B}.

Planets in orbits around more evolved stars with even lower $\log g \sim$2-3 are indeed found outside the predicted borderline, as illustrated by the 0.54~AU orbit of the BD+15~2940 planet discussed here. However, one notes that planets around giants are generally found in tighter orbits than those around sub-giants (except HD~102956~b). This suggests that, in addition to the absence of 0.1-0.6~AU companions to $M_{\star} \geq$1.2~$\Ms$ F dwarfs, another effect must influence the orbits of planets around those evolved stars. This effect has been identified as the tidal interaction between the star and the planet, which may cause an accelerated orbital decay \citep{1996ApJ...470.1187R,2009ApJ...705L..81V}. According to \cite{2009ApJ...705L..81V}, a 2~$\Ms$ star, a typical object in the observed population within the $\log g \sim$2-3 range, should have already ingested 3-5~$\Mj$ or more massive planets in orbits up to 0.2~AU before reaching $\log L_{\star}/L_{\odot}$ = 2 (or $\log g \sim$2.3) on the Red Giant Branch.

Both HD~233604~b and BD+15~2940~b fall into that part of Figure~\ref{fig-09} as well. However, HD~233604~b in 0.75~AU or of $q$ = 0.71~AU orbit around its $M_{\star}$ = 1.5~$\Ms$ host is located well within the minimum orbital radius as defined in \cite{2009ApJ...705L..81V} and \cite{2011ApJ...737...66K} for a Jupiter mass planet. With the estimated $M_{\star}$ = 1.1~$\Ms$ BD+15~2940 planet in 0.54~AU or only $q$ = 0.4~AU is located even closer, within the star's RGB tip radius \citep{2009ApJ...705L..81V}. Existence of such close-in planets suggests that the influence of star-planet tidal interactions in late stages of stellar evolution may be overestimated. The relatively large eccentricity, $e$ = 0.26 $\pm$ 0.1, of the BD+15~2940 planet also deserves more attention as one might expect its orbit to be already circularized at the distance of only 5.8 stellar radii from the star. On the other hand, it may represent a reminiscence of a recent planet-planet interaction in the system \citep{2013arXiv1301.5441V}.

Even more evolved hosts of planetary systems, with $\log g <$ 2, form a very heterogenous group with masses ranging from around 1~$\Ms$ (42~Dra) up to $M_{\star}$ = 4.5~$\Ms$ (HD~13189). As the evolution of planetary systems affected by tidal interactions and mass loss is very much stellar mass-dependent, it is difficult to discuss them together. These stars also exhibit substantial amounts of variability characterized by post-fit RV rms values easily reaching 55-60~$\mps$ \citep{2005A&A...437..743H,2007ApJ...669.1354N} which makes studying them even more difficult. 

Although the population of planetary systems detected around evolved stars has been steadily growing, it is still too meager to achieve satisfying levels of understanding of the star-planetary system interaction driven by the post-MS stellar evolution. An additional, important complication stems from the fact that luminosities of these stars are typically poorly determined, which obviously makes constraining their evolutionary status very difficult. In any case, continuing studies of such systems provide important contributions to the process of creating a general picture of the evolution of planetary systems under the influence of their evolving parent stars.

\acknowledgments
We thank Dr. Nikolai Piskunov for making SME available for us. We thank the HET resident astronomers and telescope operators for continuous support. GN, AN and MA were supported by the Polish Ministry of Science and Higher Education grant N N203 510938. GN acknowledges support from Polish Ministry of Science and Higher Education grant N N203 386237. AW was supported by the NASA grant NNX09AB36G. The HET is a joint project of the University of Texas at Austin, the Pennsylvania State University, Stanford University, Ludwig-Maximilians-Universit\"at M\"unchen, and Georg-August-Universit\"at G\"ottingen. The HET is named in honor of its principal benefactors, William P. Hobby and Robert E. Eberly. The Center for Exoplanets and Habitable Worlds is supported by the Pennsylvania State University, the Eberly College of Science, and the Pennsylvania Space Grant Consortium. This research has made extensive use of the SIMBAD database, operated at CDS (Strasbourg, France) and NASA's Astrophysics Data System Bibliographic Services. This research has made use of the The Extrasolar Planets Encyclopaedia at exoplanet.eu.
\clearpage

{}
\clearpage

\input{Table1.tex}
\input{Table2.tex}
\input{Table3.tex}
\input{Table4.tex}
\input{Table5.tex}
\input{Table6.tex}
\input{Table7.tex}
\input{Table8.tex}
\clearpage

\clearpage

\clearpage

\clearpage

\clearpage

\clearpage

\clearpage

\clearpage

\clearpage

\clearpage

\end{document}

%% file: Table1.tex
\begin{deluxetable}{llll}
\tablecaption{Stellar parameters of BD+15~2940 and HD~233604.\label{tab-01}}
\tablewidth{0pt}
\tablehead{\colhead{Parameter}                   & BD+15~2940               & HD~233604}
\startdata
$V_{T}$ (mag)                                    & 9.185 $\pm$ 0.024        & 10.406 $\pm$ 0.049\\
$(B-V)$ (mag)                                    & 1.006 $\pm$ 0.036        & 1.007 $\pm$ 0.115\\
Spectral type                                    & K0                       & K5\\
$\pi$ (mas)                                      & 1.71 $\pm$ 1.33          & --\\
$T_{eff}$ (K)                                    & 4796 $\pm$ 117           & 4791 $\pm$ 45\\
${\rm log}\,g$                                   & 2.8 $\pm$ 0.45           & 2.55 $\pm$ 0.18\\
${\rm (Fe/H)}$                                   & 0.28 $\pm$ 0.07          & -0.36 $\pm$ 0.04\\
${\rm log}\,L_{\star}/L_{\odot}$                 & 2.01 $\pm$ 0.75          & 1.75 $\pm$ 0.18\\
$M_{\star}/M_{\odot}$                            & 1.1 $\pm$ 0.2            & 1.50 $\pm$ 0.30\\
$R_{\star}/R_{\odot}$                            & 14.7 $\pm$ 2.8           & 10.9 $\pm$ 0.6\\
Age (Gyr)                                        & 7.1 $\pm$ 4.2            & 2.4 $\pm$ 1.8\\
$\vsini$ ($\kmps$)                               & 2.3 $\pm$ 0.8            & 2.0 $\pm$ 0.8\\
$\psini$ (days)                                  & 317 $\pm$ 121            & 282 $\pm$ 111\\
$K_{osc}$ ($\mps$)                               & $22^{+137}_{-19}$        & $9^{+9}_{-4}$\\
$P_{osc}$ (days)                                 & $0.68^{+0.51}_{-0.31}$   & $0.27^{+0.11}_{-0.07}$
\enddata
\end{deluxetable}

%% file: Table2.tex
\begin{deluxetable}{lrrrr}
\tablecaption{Relative radial velocities and bisector span of HD~209458.\label{tab-02}}
\tablewidth{0pt}
\tablehead{\colhead{Epoch (MJD)} & RV ($\mps$)  & $\sigma_{{\rm RV}}$ ($\mps$) & BS ($\mps$)  & $\sigma_{{\rm BS}}$ ($\mps$)}
\startdata
 53197.29482   &        -80.9   &         13.1   &         39.9   &         37.3\\
 53538.35438   &        -73.1   &         14.7   &        123.9   &         46.9\\
 53544.34874   &         49.9   &         13.9   &        -11.5   &         37.9\\
 53547.33435   &         57.4   &         14.5   &        -19.8   &         39.0\\
 53984.34746   &         65.7   &         13.3   &        -13.6   &         40.3\\
 54001.09852   &        -53.7   &         11.3   &         19.7   &         33.4\\
 54014.26729   &        -34.1   &          9.7   &          3.5   &         36.2\\
 54056.13726   &         50.6   &         12.1   &         60.2   &         42.9\\
 54448.07560   &        -59.1   &          7.3   &         33.6   &         26.2\\
 54738.09154   &         18.8   &         15.3   &         -0.0   &         52.1\\
 54756.22903   &         53.4   &         11.2   &         70.3   &         35.9\\
 54759.20674   &        -32.3   &         13.4   &          9.9   &         43.9\\
 54764.21486   &        100.9   &         10.5   &        -18.3   &         44.2\\
 54777.16241   &         27.9   &          9.5   &         38.0   &         39.3\\
 54780.15300   &        -21.4   &         12.1   &        -14.0   &         48.3\\
 54786.13373   &        -26.2   &         12.4   &         82.3   &         54.2\\
 54792.13299   &        122.2   &         10.9   &         39.0   &         44.2\\
 54813.07756   &        115.0   &         10.9   &         95.2   &         36.8\\
 55031.26789   &         63.4   &         15.3   &         18.2   &         48.1\\
 55068.16183   &        -28.8   &         15.0   &         -8.5   &         38.1\\
 55110.25281   &          3.1   &         11.8   &        -61.2   &         38.6\\
 55502.17681   &        -67.2   &         12.2   &        -19.0   &         41.8\\
 55504.18200   &         96.2   &          9.6   &        -23.4   &         26.1\\
 55506.18306   &        -60.1   &          9.5   &          9.6   &         28.4\\
 55510.15573   &        -33.5   &         11.2   &         40.4   &         39.4\\
 55510.15868   &        -35.6   &          9.2   &         39.7   &         36.5\\
 55513.16310   &        -77.0   &         12.9   &         47.1   &         36.8\\
 55515.15433   &         34.4   &         13.6   &        -47.7   &         29.6\\
 55890.13839   &        -82.0   &          7.7   &         56.8   &         25.2
\enddata
\end{deluxetable}

%% file: Table3.tex
\begin{deluxetable}{lrrrr}
\tablecaption{Relative radial velocities and bisector span of HD~88133.\label{tab-03}}
\tablewidth{0pt}
\tablehead{\colhead{Epoch (MJD)} & RV ($\mps$)  & $\sigma_{{\rm RV}}$ ($\mps$) & BS ($\mps$)  & $\sigma_{{\rm BS}}$ ($\mps$)}
\startdata
 53758.26950   &        -21.4   &          5.5   &        -28.8   &         14.2\\
 53774.21259   &         33.2   &          6.0   &        -30.2   &         14.6\\
 53835.25918   &         27.5   &          6.6   &        -27.5   &         13.4\\
 53868.17657   &        -19.8   &          6.5   &        -12.8   &         13.8\\
 54090.36895   &        -31.8   &          5.3   &        -18.4   &         17.6\\
 54090.37559   &        -30.5   &          6.6   &        -43.0   &         19.8\\
 54165.36426   &        -21.8   &          6.3   &         15.0   &         16.4\\
 54461.34054   &          8.9   &          5.6   &        -32.0   &         15.1\\
 54465.32153   &        -32.2   &          5.6   &        -13.6   &         13.5\\
 54474.31236   &         43.2   &          6.2   &        -24.9   &         16.0\\
 54767.50684   &         29.3   &          7.5   &         22.7   &         19.4\\
 54777.48514   &         12.0   &          6.2   &        -41.6   &         20.2\\
 54782.47424   &         -7.1   &          6.1   &         28.8   &         19.6\\
 54785.45165   &         18.8   &          6.5   &         -5.3   &         18.8\\
 54788.45604   &         29.1   &          5.4   &          9.4   &         16.7\\
 54858.46134   &        -28.6   &          8.1   &        -11.0   &         21.5\\
 54864.44610   &         -9.5   &          7.4   &        -12.1   &         19.1\\
 55137.50172   &        -15.8   &          5.1   &         -8.7   &         14.2\\
 55940.50047   &        -34.2   &          6.1   &          2.9   &         17.0\\
 55948.27405   &        -13.6   &          4.7   &        -60.2   &         11.1\\
 55956.24389   &         37.6   &          5.4   &         11.4   &         10.9\\
 55973.40238   &         28.1   &          6.4   &         -9.5   &         18.9\\
 55977.19226   &         12.5   &          6.3   &        -11.8   &         16.9
\enddata
\end{deluxetable}

%% file: Table4.tex
\begin{deluxetable}{lrrrr}
\tablecaption{Relative radial velocities and bisector span of HD~166435.\label{tab-04}}
\tablewidth{0pt}
\tablehead{\colhead{Epoch (MJD)} & RV ($\mps$)  & $\sigma_{{\rm RV}}$ ($\mps$) & BS ($\mps$)  & $\sigma_{{\rm BS}}$ ($\mps$)}
\startdata
 55611.51710   &        -33.2   &         10.8   &         27.7   &         56.5\\
 55612.50123   &        -33.4   &          8.4   &         -3.9   &         52.1\\
 55613.51535   &         10.5   &         15.7   &         23.3   &         52.9\\
 55615.51164   &        -24.1   &         16.8   &        -54.3   &         51.8\\
 55616.50156   &         29.1   &         10.2   &        -47.1   &         58.9\\
 55626.49751   &        -15.8   &         18.0   &         20.3   &         56.1\\
 55627.47291   &         15.2   &         10.3   &        -54.4   &         53.9\\
 55628.49111   &         12.1   &         18.9   &        -99.7   &         66.7\\
 55634.45595   &         51.8   &         10.3   &       -104.3   &         55.1\\
 55639.44095   &         -7.3   &         12.8   &         10.9   &         68.0\\
 55640.44137   &        -15.3   &         10.2   &        -77.0   &         60.8\\
 55643.41648   &        -41.3   &         16.4   &         52.0   &         61.4\\
 55644.42200   &         46.8   &         15.2   &        -91.9   &         44.8\\
 55645.40177   &         30.3   &         18.1   &        -92.1   &         63.6\\
 55801.24069   &         26.1   &         11.0   &       -154.1   &         48.4\\
 55805.21921   &         43.0   &         11.0   &       -145.8   &         44.6\\
 55814.18152   &        -28.4   &          9.0   &        -14.6   &         40.2\\
 55815.18623   &        -59.8   &         12.8   &        -32.5   &         40.5\\
 55816.18358   &         34.5   &         11.0   &        -45.3   &         43.3\\
 55817.18530   &         -1.6   &         11.5   &        -70.8   &         42.5\\
 55824.15294   &          4.7   &         11.3   &        -18.2   &         41.2\\
 55825.14860   &        -24.6   &         11.1   &        -28.3   &         42.2\\
 55826.13646   &          0.2   &         11.5   &       -129.4   &         39.5
\enddata
\end{deluxetable}

%% file: Table5.tex
\begin{deluxetable}{lll}
\tablecaption{Orbital parameters of HD~209458~b and HD~88133~b.\label{tab-05}}
\tablewidth{0pt}
\tablehead{\colhead{Parameter}              & HD~209458~b           & HD~88133~b}
\startdata
$P$~(days)                                  & 3.525 $\pm$ 0.009     & 3.415 $\pm$ 0.001\\
$T_0$~(MJD)                                 & 53199.3 $\pm$ 0.9     & 53760.7 $\pm$ 0.9\\
$K$~($\mps$)                                & 87.7 $\pm$ 5.4        & 34.2 $\pm$ 3.6\\
$e$                                         & 0.007 $\pm$ 0.060     & 0.09 $\pm$ 0.07\\
$\omega$~(deg)                              & 60 $\pm$ 108          & 15 $\pm$ 120\\
$m \sin i$~($\Mj$)                          & 0.714                 & 0.284\\
$a$~(AU)                                    & 0.0474                & 0.047\\
$RV_{0}$~($\mps$)                           & 11.1 $\pm$ 3.4        & -0.7 $\pm$ 1.4\\
$\sigma_{+}$~($\mps$)                       & 0.0                   & 0.0\\
$\rchisq$                                   & 3.13                  & 1.127\\
$\sigma_{O-C}$~($\mps$)                     & 20.54                 & 5.46\\
$\overline{\sigma_{{\rm RV}}}$~($\mps$)     & 11.87                 & 6.15\\
$\sigma_{j}$~($\mps$)                       & 16.76                 & $\leqslant$~5.46\\
$N_{obs}$                                   & 29                    & 23
\enddata
\end{deluxetable}

%% file: Table6.tex
\begin{deluxetable}{lrrrr}
\tablecaption{Relative radial velocities and bisector span of BD+15~2940.\label{tab-06}}
\tablewidth{0pt}
\tablehead{\colhead{Epoch (MJD)} & RV ($\mps$)  & $\sigma_{{\rm RV}}$ ($\mps$) & BS ($\mps$)  & $\sigma_{{\rm BS}}$ ($\mps$)}
\startdata
 53452.36375   &        -46.8   &          7.8   &         15.9   &         17.0\\
 53499.41176   &          8.4   &          7.5   &          5.9   &         17.8\\
 53844.29702   &        -40.7   &          5.5   &         -4.5   &         14.5\\
 53866.40534   &         10.7   &          6.6   &         -7.3   &         18.5\\
 53875.39832   &         19.5   &          5.6   &         -4.7   &         13.0\\
 53892.16699   &         37.0   &          6.3   &         25.4   &         16.6\\
 53902.31759   &         62.6   &          5.5   &          1.7   &         14.9\\
 53910.29146   &         23.5   &          5.9   &          0.3   &         14.4\\
 53955.16666   &        -20.7   &          5.2   &        -45.9   &         16.1\\
 53971.13068   &        -40.3   &          5.9   &         -8.1   &         14.8\\
 54215.25852   &        -32.1   &          6.6   &         28.2   &         17.0\\
 54225.42513   &        -27.4   &          7.9   &        -15.1   &         26.1\\
 54248.35726   &        -29.3   &          6.1   &        -36.4   &         17.1\\
 54276.28661   &         -6.9   &          7.6   &         42.4   &         19.4\\
 54290.25121   &         38.6   &          7.3   &        -45.6   &         19.5\\
 54320.17812   &         37.9   &          7.6   &          4.6   &         20.9\\
 54515.45514   &        -30.1   &          7.6   &        -30.5   &         19.2\\
 54564.30494   &         26.3   &          7.8   &        -32.5   &         23.3\\
 54595.22798   &         47.0   &          7.7   &          9.9   &         17.4\\
 54624.34067   &        -38.2   &          7.2   &        -24.2   &         16.3\\
 54692.15651   &         17.6   &          8.2   &         25.1   &         19.1\\
 55018.25340   &         22.0   &          8.7   &        -17.2   &         17.1\\
 55049.18227   &        -54.8   &          6.4   &          9.2   &         15.0\\
 55254.40663   &         28.6   &          7.8   &          7.2   &         20.8\\
 55280.37572   &         27.5   &          8.0   &         34.5   &         22.8\\
 55582.52001   &         13.9   &          7.6   &        -11.6   &         20.7\\
 55621.40203   &         -1.5   &          7.5   &         30.7   &         24.0\\
 55621.41021   &        -12.4   &          6.0   &         16.7   &         15.0\\
 55674.46852   &          6.9   &          7.5   &         -5.5   &         18.4\\
 55719.34138   &         -4.3   &          6.6   &        -29.0   &         15.6\\
 55742.28347   &        -16.6   &          6.9   &        -25.8   &         15.5\\
 55750.25899   &        -37.5   &          7.7   &          4.4   &         14.6\\
 55763.22678   &        -53.9   &          8.5   &        -37.6   &         17.2\\
 55776.18611   &        -68.1   &          6.6   &        -11.2   &         16.1\\
 55795.13382   &         -3.2   &          6.4   &        -11.2   &         15.1\\
 55801.11490   &         35.3   &          6.1   &          3.0   &         15.4\\
 55808.08974   &         60.0   &          6.7   &        -17.4   &         17.1\\
 55814.09263   &         43.2   &          6.3   &        -18.1   &         17.5
\enddata
\end{deluxetable}

%% file: Table7.tex
\begin{deluxetable}{lrrrr}
\tablecaption{Relative radial velocities and bisector span of HD~233604.\label{tab-07}}
\tablewidth{0pt}
\tablehead{\colhead{Epoch (MJD)} & RV ($\mps$)  & $\sigma_{{\rm RV}}$ ($\mps$) & BS ($\mps$)  & $\sigma_{{\rm BS}}$ ($\mps$)}
\startdata
 53015.25986   &        -28.9   &         10.6   &          4.8   &         26.2\\
 53360.29494   &       -231.5   &         10.7   &        -57.9   &         22.0\\
 54064.39510   &         24.8   &          7.9   &         43.7   &         19.3\\
 54136.38069   &       -172.2   &         10.6   &         -2.6   &         20.6\\
 54208.19384   &        142.1   &          9.2   &         12.1   &         18.5\\
 54399.46353   &        135.1   &          7.5   &          8.3   &         20.6\\
 54420.41574   &        138.1   &          8.3   &          5.2   &         24.2\\
 54500.17910   &       -193.3   &         10.0   &        -50.3   &         25.1\\
 54564.21884   &         80.0   &          9.0   &         29.2   &         21.4\\
 54604.14329   &        119.2   &          8.5   &         12.1   &         20.5\\
 54763.46178   &         83.8   &          9.5   &        -20.0   &         26.1\\
 54782.43096   &        193.9   &          9.0   &         14.1   &         28.3\\
 54839.45575   &         -7.2   &         10.1   &          0.1   &         18.1\\
 54865.19007   &       -134.9   &         10.6   &         15.4   &         26.7\\
 54891.33472   &       -183.9   &         11.9   &         -7.2   &         25.9\\
 55173.35355   &        165.5   &          9.9   &         22.4   &         26.1\\
 55173.37553   &        159.9   &          8.6   &          1.3   &         24.0\\
 55176.52706   &        181.0   &          9.5   &         64.1   &         28.6\\
 55214.26133   &          5.7   &          9.9   &         72.5   &         25.0\\
 55223.20532   &        -28.9   &          8.5   &         36.4   &         24.8\\
 55254.36068   &       -151.2   &         10.3   &        -43.7   &         24.1\\
 55307.19447   &       -102.0   &          8.1   &        -10.7   &         16.1\\
 55486.49435   &       -145.8   &          7.5   &         13.2   &         21.5\\
 55487.49885   &       -155.4   &          6.9   &         -8.3   &         18.0\\
 55502.47794   &        -56.7   &          9.7   &         44.5   &         26.9\\
 55519.38507   &         18.2   &         10.6   &         35.5   &         30.3\\
 55520.39557   &         35.2   &          8.9   &        -10.1   &         17.4\\
 55645.26266   &       -155.3   &          9.1   &         49.6   &         23.1\\
 55659.21377   &       -190.8   &          8.2   &         26.4   &         15.6\\
 55671.19521   &       -185.3   &          7.2   &          5.7   &         17.3\\
 55956.18941   &        167.2   &          6.9   &         34.3   &         17.1\\
 55964.38336   &        152.6   &          7.2   &          3.6   &         16.4\\
 55971.16102   &         59.5   &          7.5   &        -28.4   &         15.9
\enddata
\end{deluxetable}

%% file: Table8.tex
\begin{deluxetable}{lll}
\tablecaption{Orbital parameters of BD+15~2940~b and HD~233604~b.\label{tab-08}}
\tablewidth{0pt}
\tablehead{\colhead{Parameter}              & BD+15~2940~b           & HD~233604~b}
\startdata
$P$~(days)                                  & 137.48 $\pm$ 0.34      & 192.00 $\pm$ 0.22\\  
$T_0$~(MJD)                                 & 53464 $\pm$ 16         & 53022 $\pm$ 77\\     
$K$~($\mps$)                                & 42.7 $\pm$ 4.4         & 177.8 $\pm$ 4.3\\    
$e$                                         & 0.26 $\pm$ 0.10        & 0.05 $\pm$ 0.03\\    
$\omega$~(deg)                              & 302 $\pm$ 33           & 281 $\pm$ 43\\       
$m \sin i$~($\Mj$)                          & 1.11                   & 6.575\\              
$a$~(AU)                                    & 0.539                  & 0.747\\              
$RV_{0}$~($\mps$)                           & -2.5 $\pm$ 3.3         & -16.9 $\pm$ 3.7\\    
$\sigma_{+}$~($\mps$)                       & 18.0                   & 9.0\\                
$\rchisq$                                   & 0.989                  & 2.76\\               
$\sigma_{O-C}$~($\mps$)                     & 17.89                  & 19.13\\              
$\overline{\sigma_{{\rm RV}}}$~($\mps$)     & 6.97                   & 9.04\\               
$\sigma_{j}$~($\mps$)                       & 16.48                  & 16.86\\              
$N_{obs}$                                   & 38                     & 33                   
\enddata         
\end{deluxetable}